%% file: evlog_ieee_issre.tex
\documentclass[conference]{IEEEtran}

\AtBeginDocument{%
  \providecommand\BibTeX{{%
    \normalfont B\kern-0.5em{\scshape i\kern-0.25em b}\kern-0.8em\TeX}}}

\def\name{EvLog}
\usepackage{algorithm}  
\usepackage{algpseudocode} 
\usepackage{enumitem}
\usepackage{graphbox}
\usepackage{amsmath, amssymb}
\usepackage{balance}

\usepackage[TABBOTCAP]{subfigure}
\usepackage{booktabs}
\usepackage{multirow}
\usepackage[]{collab}
\usepackage{tcolorbox}
\usepackage{colortbl}

\usepackage{colortbl}

\definecolor{maroon}{cmyk}{0,0.87,0.68,0.32}

\definecolor{background}{rgb}{0.94, 0.97, 1.0}
\definecolor{edge}{rgb}{0.32, 0.48, 0.72}

\begin{document}
\title{EvLog: Identifying Anomalous Logs over Software Evolution}

\author{
  \IEEEauthorblockN{
    Yintong Huo\IEEEauthorrefmark{2},
    Cheryl Lee\IEEEauthorrefmark{2},
    Yuxin Su\IEEEauthorrefmark{3}\IEEEauthorrefmark{1}, 
    Shiwen Shan\IEEEauthorrefmark{3},
    Jinyang Liu\IEEEauthorrefmark{2}, and
    Michael R. Lyu\IEEEauthorrefmark{2}
  }


  \IEEEauthorblockA{\IEEEauthorrefmark{2}Department of Computer Science and Engineering, The Chinese University of Hong Kong, Hong Kong, China.\\
    Email: \{ythuo, jyliu, lyu\}@cse.cuhk.edu.hk, cheryllee@link.cuhk.edu.hk}

  \IEEEauthorblockA{\IEEEauthorrefmark{3}Sun Yat-sen University, Zhuhai, China.
    Email: suyx35@mail.sysu.edu.cn, shanshw@mail2.sysu.edu.cn}
}

\maketitle

\input{sections/0Abstract}

\input{sections/1Introduction}
\input{sections/2EmpiricalStudy}
\input{sections/3Problem}

\input{sections/4Approach}

\input{sections/5Implementation}
\input{sections/6Experiment}

\input{sections/7CaseStudy}
\input{sections/8ThreatLiterature}

\input{sections/9Conclusion}

\balance
\normalem
\bibliographystyle{IEEEtran}
\bibliography{references}
\end{document}

%% file: sections/0Abstract.tex
\begin{abstract}
Software logs record system activities, aiding maintainers in identifying the underlying causes for failures and enabling prompt mitigation actions. However, maintainers need to inspect a large volume of daily logs to identify the anomalous logs that reveal failure details for further diagnosis.
Thus, how to automatically distinguish these anomalous logs from normal logs becomes a critical problem.
Existing approaches alleviate the burden on software maintainers, but they are built upon an improper yet critical assumption: logging statements in the software remain unchanged. While software keeps evolving, our empirical study finds that evolving software brings three challenges: log parsing errors, evolving log events, and unstable log sequences.
In this paper, we propose a novel unsupervised approach named Evolving Log analyzer (\name) to mitigate these challenges.
We first build a multi-level representation extractor to process logs without parsing to prevent errors from the parser. 
The multi-level representations preserve the essential semantics of logs while leaving out insignificant changes in evolving events.
\name~then implements an anomaly discriminator with an attention mechanism to identify the anomalous logs and avoid the issue brought by the unstable sequence.
\name~has shown effectiveness in two real-world system evolution log datasets with an average F1 score of 0.955 and 0.847 in the intra-version setting and inter-version setting, respectively, which outperforms other state-of-the-art approaches by a wide margin.
To our best knowledge, this is the first study on localizing anomalous logs over software evolution.
We believe our work sheds new light on the impact of software evolution with the corresponding solutions for the log analysis community.
\let\thefootnote\relax\footnotetext{$^{*}$ Corresponding author.}
\end{abstract}

%% file: sections/1Introduction.tex
\section{Introduction}

Nowadays, intelligent log analytics is designed to manage overwhelming logs~\cite{shilpika2019mela} for failure troubleshooting, and anomaly detection~\cite{liu2023scalable}.
Existing automated log analytics can be categorized into two types based on granularity: coarse-grained tasks and fine-grained tasks.
Coarse-grained models, such as the anomaly detectors~\cite{deeplog2017} and failure predictors~\cite{klinkenberg2017data}, detect (or predict) anomalies given the logs from a period of time.
Taking anomaly detection as an example, the model accepts a session of logs to determine whether an anomaly exists in this session. 
Although the coarse-grained models show promising results in open datasets, they provide limited evidence of failure diagnosis for software maintainers. 
On the other hand, fine-grained tasks aim to further identify the individual/single anomalous logs within a session showing possible interpretations of the failure~\cite{zhang2021onion, jia2017logsed, logan2016}. 
Even if coarse-grained models free maintainers from inspecting massive log lines, it is still time-consuming to analyze hundreds of log lines within a session to find the anomalous log for troubleshooting~\cite{meng2021logclass}. To ease the burden of software maintainers, we focus on this more challenging yet significant task, individual anomalous log identification, in this paper.


An \textit{anomalous log} signals an anomaly in the system, such as network error~\cite{meng2021logclass}. The following example shows a log message that may indicate a connection problem caused by a network fault within the system:
\begin{center}
\small
    \fbox{
    Container launch failed for container\_32h: Connection refused.
    }
\end{center}
Anomalous logs are crucial for diagnosing failures, but they are often accompanied by numerous normal logs, which can be overwhelming for maintainers.
To distinguish them from normal run-time logs,
existing studies~\cite{jia2017logsed, logan2016, amar2018using, wang2021groot} constructed a \textit{reference model} from training log sequences and then identified which log violated the reference model. Specifically, they abstracted log event sequences into a directed graph via either a finite state machine (FSM)~\cite{jia2017logsed, logan2016, amar2018using} or causal dependencies~\cite{wang2021groot} as the reference model. Subsequently, any deviations from this model would be regarded as an indication for anomaly and marked for troubleshooting.

However, both FSM-based and causal graph-based approaches following the \textit{closed-world assumption} suffer limitations for processing the unseen data.
However, after the initial version is released, software experiences continual development to fulfill customers' demand, to fix bugs, and to extend to new functionalities, which is well-known as \textit{software evolution}~\cite{lehman2002software, shang2014exploratory}. 
Previous studies pointed out that logging statements change over software evolution is so pervasive that around 33\% of the log are revised as after-thoughts~\cite{yuan2012characterizing, chen2017characterizing}.
The changed logging statements during the evolution activities raise challenges for existing approaches:

\textbf{\textit{(1) Parsing errors.}} Log parsers extract static events (e.g., \textit{Container launch failed for $<$*$>$: Connection refused.}) and dynamic parameters (e.g., \textit{container\_32h}) from log messages. However, as discussed in Section~\ref{sec:empirical-parsing-errors}, parsers may misalign revised log events in evolving software versions, causing log parsing errors. These parsing errors further downgrade the subsequent log analytics performance. 
\textbf{\textit{(2) Evolving events.}} Even if state-of-the-art parsers work as expected, software evolution brings new logging statements or paraphrases old logging statements, which we refer as \textit{evolving events} in this paper. 
\textbf{\textit{(3) Unstable sequences.}} Apart from log events, the log sequences from running identical jobs can vary, named \textit{unstable sequences}. Such variation can be caused by interleaving logs produced from multiple threads~\cite{jia2017logsed}. Moreover, software evolution may alter the function invocation sequences, leading to new sequential patterns. 



\begin{figure}[tbp]
    \centering 
    \includegraphics[width=0.45\textwidth]{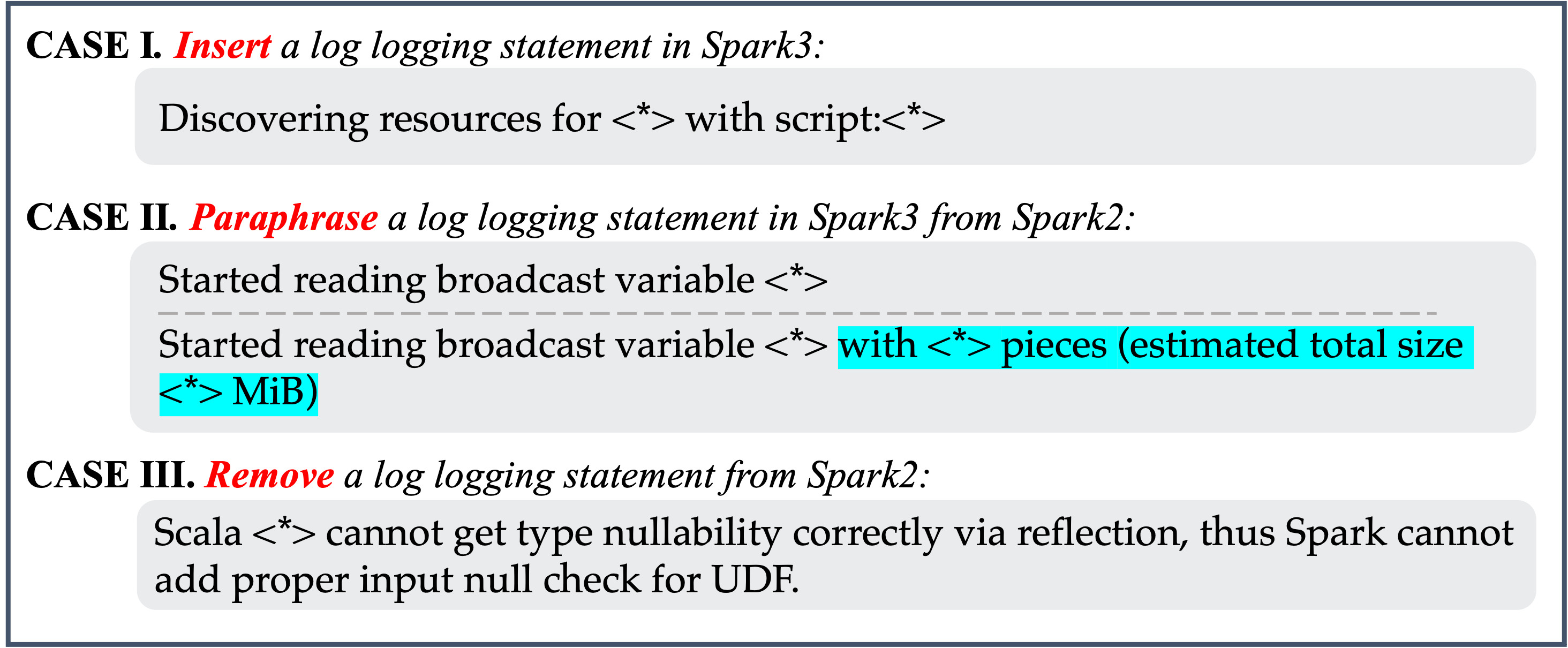} 
        \vspace{-0.1in}
    \caption{Evolving logging statement cases for Spark2 and Spark3.} 
    \vspace{-0.2in}
    \label{fig:empirical-case} 
\end{figure}

While solutions to the first two challenges still remain unexplored, there have been several attempts to handle the third challenge. For example, previous study~\cite{jia2017approach} tried to resolve the interleaving logs by considering multiple predecessors and successors of a log event, instead of just the direct ones. Another study~\cite{fu2014digging} mitigated unstable sequences challenge by learning causal relationships between event pairs from historical data. Nevertheless, none of the existing approaches considered the software evolution scenario, which can negatively impact the performance of identifying anomalous logs if left unaddressed.



To address the above challenges, we propose an unsupervised anomalous log identification solution over software evolution. 
The design of our approach is based on two insights: \textit{1) the majority of logs are normal in a healthy system; and 2) the anomalous logs are unknown a priori because we cannot iteratively inject all kinds of failures.} 
In particular, we design~\name~with two steps.
The first step aims to tackle the \textit{parsing errors} and the \textit{evolving events} issues. We derive multi-level representations directly from logs to prevent introducing \textit{parsing errors}. The representations at different levels undertake different functions: 1) the semantic-rich representation aims to fully retain semantics from log messages, which is extracted by pre-trained language models; and
2) the abstract representation to align similar logs across software evolution, which is derived from the hierarchical clustering approach. Such multi-level representations maintain the pertinent semantics while leaving out unnecessary trifles to address the \textit{evolving events} issue.

In the second step, we address the \textit{unstable sequence} issue by constructing an anomaly discriminator with an attention mechanism. 
The core idea is to learn a transformation function (e.g., neural networks) that embeds normal log features (source domain) to stay close (enclosed in a hyper-sphere) to a target domain, then the logs that are largely distant from this hyper-sphere are considered as anomalous ones. 
Specifically, \name~constructs log features for each single log and its surrounding log contexts based on multi-level representations. It then applies neural networks to discriminate the anomalous logs instead of rigorously comparing new sequences with existing ones.
Once trained, \name~can be directly applied to a future software version without any fine-tuning. 


Our new approach is evaluated using two realistic datasets (i.e., \textsc{LogEvol}) and a synthetic dataset (i.e., \textsc{SynEvol}) to simulate logging evolution.
The experiment results illustrate that \name~reaches a promising average F1 score of 0.955 and 0.847 in intra-version identification and inter-version anomalous log identification on two representative system logs, respectively. 

\begin{table}[tbp]
\footnotesize
    \centering
    \caption{Logging evolution ratio between Spark2 and Spark3.}
        \vspace{-0.1in}
    \begin{tabular}{l||c|c|c|c}
    \toprule
   Percentage & Unchanged & Inserted & Paraphrased  & Removed \\
    \midrule
   Log message & 91.16\% & 0.07\% & 8.75\%  & 0.02\% \\
   Logging statement & 76.12\% & 12.69\% & 1.49\%  & 9.70\% \\
         \bottomrule
    \end{tabular}
    \label{tab:empirical-ratio}
            \vspace{-0.2in}
\end{table}
 
To conclude, the contribution of this paper is threefold:
\begin{itemize}[leftmargin=*, topsep=0pt]
    \item We empirically identify three challenges (i.e., parsing errors, evolving events, unstable sequence) brought by software evolution for anomalous log identification, which has never been properly addressed before.
    \item To overcome the above challenges, we develop \name, an unsupervised anomalous log identification approach with a multi-level representation extractor and an anomaly discriminator. To our best knowledge, \name~is the first solution to tackle the problem of identifying anomalous logs over software evolution.
    \item By evaluating \name~on real system log datasets and a synthetic dataset, we show our approach can effectively identify anomalous logs across different software versions without fine-tuning or manual labeling. Artifacts are released for research purposes at https://github.com/YintongHuo/EvLog.
\end{itemize}

%% file: sections/2EmpiricalStudy.tex
\section{Motivating study}\label{sec:empirical}

\subsection{How do logging statements evolve?}
Developers may modify logging statements when updating the software, producing unseen log messages in system run-time for maintenance.
To examine how logging statements evolve during software updates, we analyze Spark, an open-source cluster computing system for the parallel processing of large-scale data.
In particular, we run benchmark workloads in Spark 2.4.0 (denoted as Spark2) and Spark 3.0.3 (denoted as Spark3) with details shown in Section~\ref{subsec:datacollection} and compare the collected log messages.

We categorize the change of logging statements into three types: \textit{insert}, \textit{paraphrase} and \textit{remove}. We show three cases in Fig.~\ref{fig:empirical-case}, where ``$<$*$>$'' refers to the dynamic parameters generated in running time. 
In Case I, a new logging statement is added in Spark3 to indicate the attached resources.
In Case II, the logging statement is paraphrased by adding information on the number of pieces and the estimated size of the variable to gain a deeper understanding of the system performance.
In Case III, a logging statement is removed from Spark2 due to the deprecation of "\textit{UserDefinedFunction}" in Spark3.


\begin{figure*}[tbp]
    \centering 
        \vspace{-0.1in}
    \includegraphics[width=\textwidth]{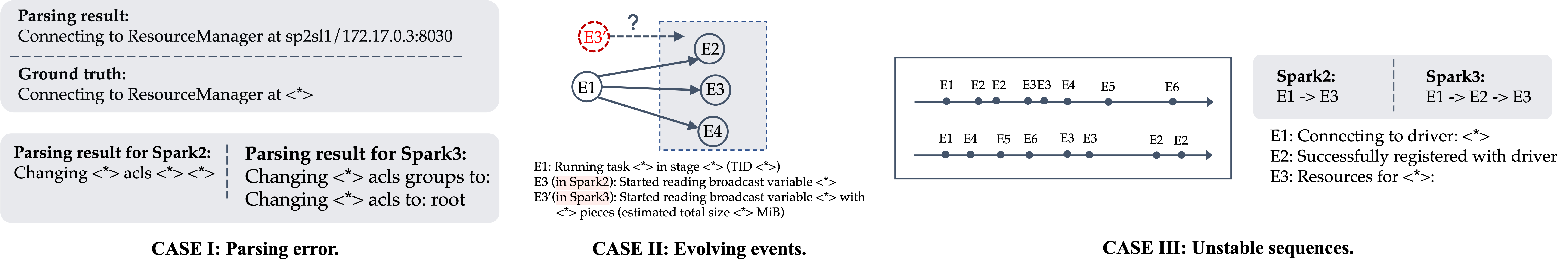} 
    \vspace{-0.3in}
    \caption{Three challenges brought by software evolution. 
    E1, E2, etc., represent different log events. (1) The parsing error case shows the incorrectly parsed log messages that will impact the subsequent log analysis, (2) The evolving event case exhibits that a paraphrased logging statement will mislead the reference model, and (3) The unstable sequence case depicts how a new logging statement E2 can alter produced log sequences.} 
    \label{fig:empirical-affects} 
    \vspace{-0.2in}
\end{figure*}

Table~\ref{tab:empirical-ratio} displays the statistics of the three types of changes on collected log messages and logging statements. 
It is observed that nearly 24\% logging statements are changed from Spark2 to Spark3, resulting in almost 10\% changed log messages. Although 12.69\% and 9.70\% logging statements are inserted or removed, respectively, they only make up less than 0.1\% collected logs, meaning they appear in a low frequency. However, the high proportion of paraphrased logs implies developers are likely to modify the commonly-used logging statements.
To conclude, logging statements change over software evolution.
The non-negligible amount of changes motivates us to reckon with the software evolution issue.

%


\subsection{How does evolution raise challenges for anomalous log identification approaches?}

 \subsubsection{Parsing errors}\label{sec:empirical-parsing-errors}

Log parsers extract constant strings (i.e., events) and run-time parameters from log messages. However, existing log identification models \textit{only} use the extracted events and do not consider the original log messages. This can be problematic because log parsers can introduce errors, and the evolution of logs over time can make parsing even more challenging~\cite{wang2022spine}. 
CASE I in Fig.~\ref{fig:empirical-affects} displays two parsing mistakes from a widely-used parser, Drain~\cite{he2017drain}, where $<$*$>$ denotes parameters. The top one is caused by confusing parameters with constant strings, and the bottom one shows inconsistent parsing results in Spark2 and Spark3. Since current log parsers are parameter-sensitive and not versatile enough~\cite{chu2021prefix}, and the hyper-parameters that work well for one software version may not be suitable for others. Since systematic log analytics should operate on raw log messages, it is essential to find ways to avoid parsing mistakes.



\subsubsection{Evolving events}
Log identification models also face a challenge in dealing with evolving events. Typically, these models detect anomalous logs by examining whether the actual next log is in line with the predicted next logs based on contextual information.
The idea works well when all the events are known; however, if the actual next log is an unseen event, it can never be matched with any predicted next logs. For instance, in CASE II of Fig.~\ref{fig:empirical-affects}, event E3' cannot be matched with the predicted logs since it is a paraphrase of event E3.
According to its decision logic, such inconsistency leads to a significant issue where all unseen events are treated as false positives. 
This issue becomes severe for all existing log-based approaches considering that 8.75\% of the collected logs have been paraphrased.

\subsubsection{Unstable sequences}

Ideally, we expect that the log message sequences perfectly match the execution sequences of a program. 
However, there are situations where log messages from different threads can interleave, resulting in what we refer to as "unstable sequences". 
Additionally, introducing new logging statements in a software update can create new log events during run-time, leading to sequential pattern changes.
As shown in CASE III of Figure~\ref{fig:empirical-affects}, unstable sequences can be caused by interleaving logs~\cite{jia2017logsed} and new log events from software evolution. To resolve the issue, identifying relevant and informative log messages in a sequence is of great essence.

In summary, our empirical findings suggest that logging evolution can affect existing models in three ways: the potential parsing errors, the evolving events, as well as the unstable sequences.
These influential factors have never been explored, yet their impact can hardly be ignored.



%% file: sections/3Problem.tex
\section{Problem illustration}\label{sec:problem}


In this paper, we consider the anomalous log identification problem as in the literature~\cite{jia2017logsed, logan2016,tak2016logan}, which enables pinpoint a collection of fault-indicating anomalous logs~\cite{meng2021logclass}.
Given a sequence of log messages $s=s_1,s_2,...,s_n$, the task asks the model to find a set of anomalous logs $AL=\{s_i| 1\leq i\leq n$\} within the message sequence.
Compared with the anomaly detection task that determines whether a problem exists in a session (session-based), anomalous log identification is a more fine-grained and challenging task that needs to localize individual fault-indicating logs (message-based). We use \textit{context} to represent the surrounding logs of a specific log (named as the \textit{center log}) and analyze whether the log is anomalous based on its context. 

\begin{figure}[tbp]
    \centering 
    \includegraphics[width=0.48\textwidth]{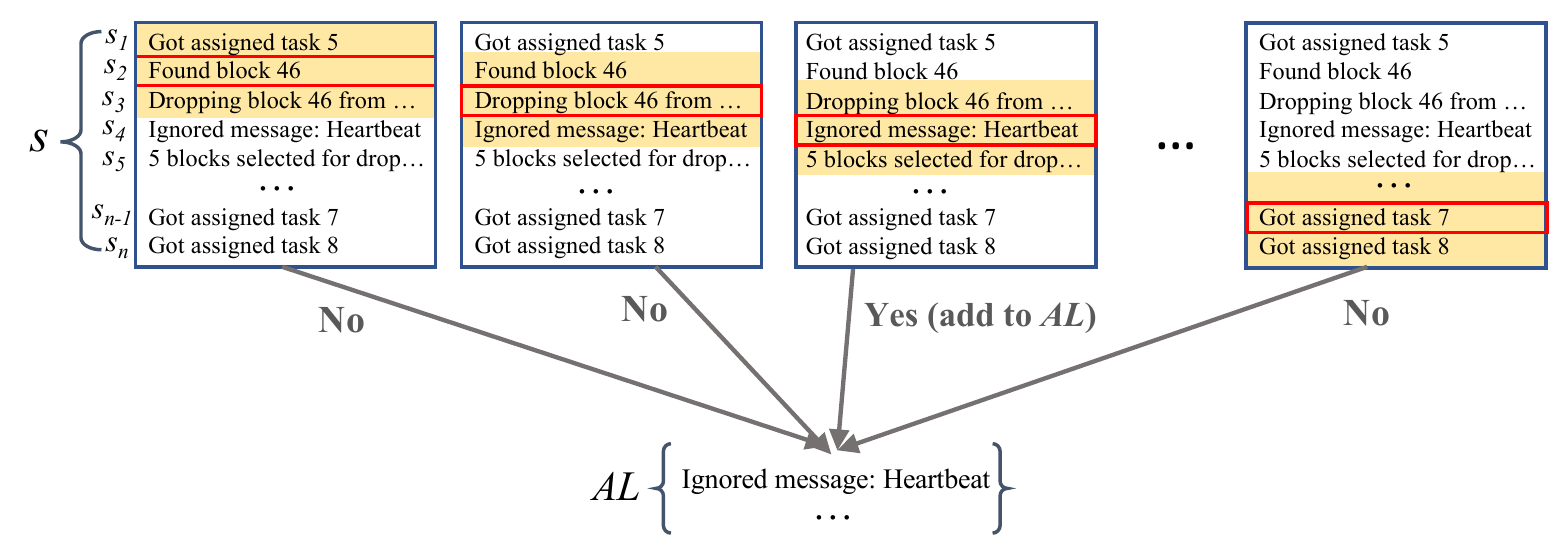} 
    \vspace{-0.1in}
    \caption{Anomalous logs localization problem illustration.} 
        \vspace{-0.2in}
    \label{fig:problem} 
\end{figure}

\begin{figure*}[tbp]
    \centering 
    \vspace{-0.1in}
    \includegraphics[width=\textwidth]{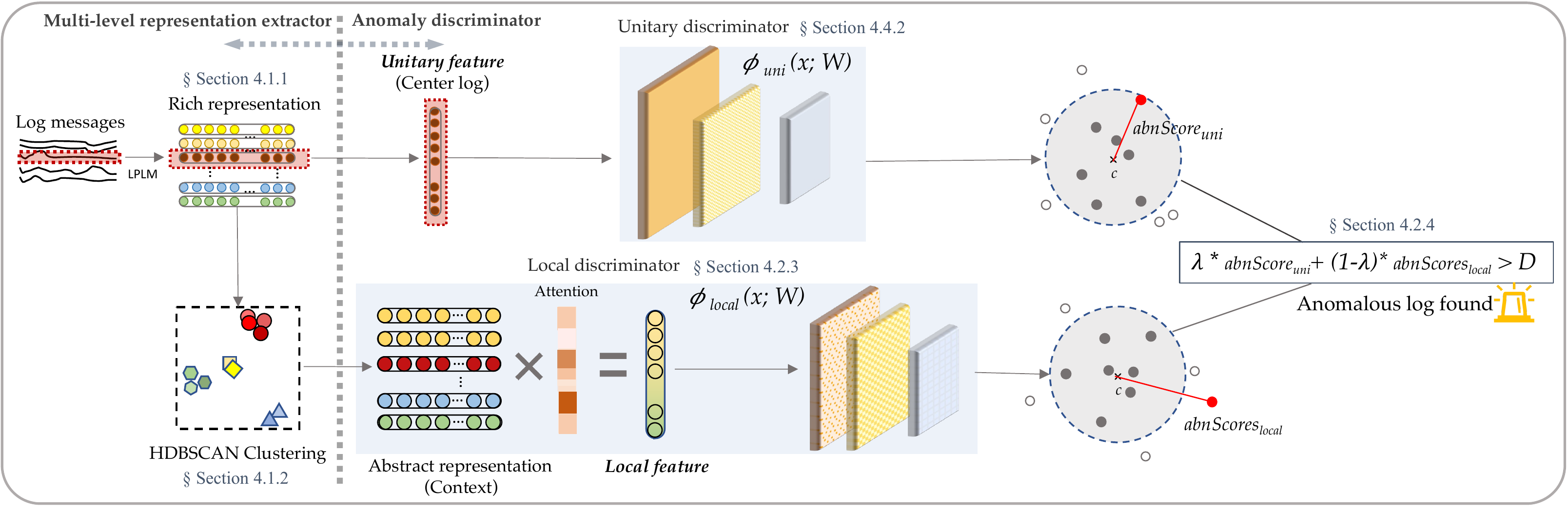} 
    \vspace{-0.25in}
    \caption{\name~with a multi-level representation extractor and an anomaly discriminator.} 
    \vspace{-0.2in}
    \label{fig:model} 
\end{figure*}

To resolve the subset $AL$, we check every individual log, that is, for all $s_i \in s$, the model determines whether the center log $s_i$ is an anomalous log in the given context of $s_i$. We will add the center log into $AL$ if it is considered anomalous. Fig.~\ref{fig:problem} shows the identification process, where the center log and its corresponding context are highlighted with a red rectangular and yellow background, respectively.

%% file: sections/4Approach.tex
\section{Approach}\label{sec:approach}


This section introduces our novel approach, called \name, shown in Fig.~\ref{fig:model}, in tackling the anomalous log identification challenges over software evolution. \name~has two components, i.e., a \textit{multi-level representation extractor} to derive multi-level robust log representations, followed by an \textit{anomaly discriminator} with the attention mechanism to pinpoint the anomalous logs. 
In particular, the multi-level representation extractor targets at extracting \textit{rich representations} as informative as possible and \textit{abstract representations} to capture high-level commonalities among similar logs.
Then these representations are fed into the anomaly discriminator to automatically localize the anomalous logs in an unsupervised manner.

\subsection{Multi-level representation extractor}
We exploit multi-level representations with various information from log messages to semantically understand them.
This section illustrates how to extract multi-level semantic representations, that is, a \textit{rich representation} and an \textit{abstract representation}. 
The low-level rich representation provides a concrete understanding of a certain log.
In contrast, the high-level abstract representation captures the commonality of logs with similar semantics, regardless of their slight differences (e.g., parameters difference, revised log events).


\subsubsection{Rich representation}
Semantics in both log events and their corresponding parameters has advantageous for log analysis~\cite{huo2021semparser, huo2022logvm}.
To obtain informative representations from logs with respect to their semantics, we fine-tune a pre-trained language model~\cite{devlin2018bert} (PLM) on our collected log datasets. 

The PLMs have shown the powerful semantic encoding ability for many software engineering tasks, such as log-based anomaly detection~\cite{le2021log} and code comprehension~\cite{chirkova2021empirical}. In our work, to overcome potential parsing errors and to make the best usage of information inside log messages, \name~acquires domain-specific semantic representations via PLMs.
On one side, system logs share some fundamental knowledge with natural languages since humans write logging statements. After being trained on a large corpus, the PLMs learn more information about word senses, not limited to system logs. 
On the other side, we notice that these PLMs are not sufficient for domain-specific tasks due to the knowledge gap. Hence, we fine-tune the massive language models to further capture domain-specific semantics. In specific, we employ the widely-used masked language modeling strategy~\cite{mastropaolo2022using, niu2022spt, izadi2022codefill} to fine-tune the PLMs, by randomly masking 10\% tokens in each log and asking the model to predict the masked tokens.

Specifically, given a log message $x$, the rich representation $x_{rich}$ is designed to capture its detailed semantics. This parser-free representation extractor accepts log messages instead of events, allowing it to get away from potential parsing mistakes.
\vspace{-15pt}
\begin{algorithm}
\footnotesize
  \caption{Abstract representation acquisition.}  
  \label{algo:cluster}
  \begin{algorithmic}[1]  
    \Require Rich representation to be clustered $E=[e_1,e_2,...,e_n]$
  \Ensure Abstract representation $C=[c_1,c_2,...,c_n]$

\State $C=[]$
\State Centroid=\{\} 
\State $E'$ = \textsc{PrincipalComponentAnalysis}($E$)
\State ClusterIds = \textsc{HDBSCAN}($E'$)


\State \textit{"""Compute centroid for each cluster"""}
\ForAll {ClusterID from 1 $\to$ \textsc{Set}(ClusterIds)}
    \State Centroid[ClusterID] = \textsc{Mean}(E'[ClusterIds==ClusterID])
\EndFor

\State \textit{"""Compute abstract representation"""}
\ForAll {$i$ from 1 $\to$ n}
\State $C$\textsc{.append}(Centroid[ClusterIds[$i$]])
\EndFor
  \end{algorithmic}  
\end{algorithm}  
\vspace{-15pt}

\subsubsection{Abstract representation}

Apart from the rich representation, we also extract a high-level semantic representation, $x_{abs}$, that remains stable on similar log events over logging evolution.
To this end, we develop a cluster-based approach on top of the rich representations. 
Previous studies~\cite{li2020clustering, silic2013prediction} have demonstrated the effectiveness of clustering approaches in grouping similar texts together based on their intrinsic characteristics. Motivated by theirs, we also employ a cluster-based approach to group log messages. 
Specifically, we adopt the idea from the previous log clustering study~\cite{yang2021semi} using Hierarchical Density-Based Spatial Clustering of Applications with Noise (HDBSCAN)~\cite{mcinnes2017accelerated}, whose efficiency and effectiveness has been presented in many domains~\cite{cifariello2019wiser, chen2020practical}. 
Compared to other clustering approaches, HDBSCAN inherits two special advantages for our scenario: (1) It can automatically extract the ``dense" cluster without pre-defining the number of clusters (e.g., Kmeans~\cite{na2010research}), which is important in the case that we may never know the number of clusters of logs. (2) HDBSCAN has a few parameter numbers, and its robustness to parameter choice~\cite{yang2021semi, mcinnes2017accelerated} makes it versatile for diverse log data.


Eventually, the abstract representation $x_{abs}$ for each log message $x$ is the centroid of its corresponding cluster by averaging all points (logs) belonging to the cluster. Algorithm~\ref{algo:cluster} shows the abstract representation computation process.

\subsection{Anomaly discriminator}
This section illustrates how we pinpoint anomalous logs from the acquired rich and abstract representations. 
In particular, for each input log, the unitary discriminator processes the log, and the local discriminator processes the log's context. Two processed results are integrated as the final output.

\subsubsection{Basic idea} 
Existing unsupervised log-based  reference sequences models~\cite{ deeplog2017,jia2017logsed,meng2019loganomaly} build a reference model from training data and check whether the testing log violates the prediction from the model. Unfortunately, these models are ineffective at handling evolving events over software evolution. Moreover, the anomalous logs are unknown a priori because we cannot iteratively inject all types of faults. Thus, we propose a different approach to handle the problem by learning the ``normality" of the normal log features instead of predicting the subsequent events.

Motivated by the Support Vector Machine~\cite{noble2006support} (SVM) that learns a hyperplane to separate data, our idea is to develop a neural network that learns a hyper-sphere to separate normal logs and anomalous logs.
The neural network maps log features (in the source domain) to a target domain where normal features stay as close as possible (enclosed in a hyper-sphere). We measure the distance between a mapped log feature and the center of the hyper-sphere as \textit{normality} (e.g., grey circle in Figure~\ref{fig:model}), with logs far from the center being considered anomalous due to deviating from the normality.
In this way, logs with evolving events can be transformed into the target domain where they are close to the previous semantically similar logs, minimizing adverse effects on the anomaly discriminator results. For example, if a normal log is paraphrased during software evolution, the evolved log with similar semantics will be mapped within the normality.
This new approach in localizing anomalous logs is superior via two advantages: 
1) It delivers better performances than other traditional methods due to the neural network's proven learning ability. 2) Our approach frees humans from labor-intensive labeling since it can learn the normality naturally in an unsupervised manner from large-scale normal logs that can be easily collected from stable software.
Specifically, the goal is to train a neural network model (mapped features from the source domain to the target domain) while minimizing the hyper-sphere volume that encloses the normal data features in the target domain. In this way, the model is forced to learn implicit semantics since it must map the normal log features closely to the hyper-sphere's center.
Thus the unseen log events with similar semantics can also be embedded close in the target domain.
To achieve the above goal, the objective function is:
\begin{equation}
    \begin{aligned}
    J = \mathop{\min}_{W} \ \ \frac{1}{n}\sum_{i=1}^{n}\| \phi(x_i; W) - c \|^2 + \frac{\alpha}{2} \|W\|^2,
    \end{aligned}
    \label{eq:objctive-func}
\end{equation}
where $\phi(x_i; W)$ refers to using the model $\phi$ with its parameters $W$ to map each input sample $x_i \in x$ to a hyperspace $\mathbb{R}^n$; $c \in \mathbb{R}^n$ refers to the hyper-sphere center; the last term serves as a regularization term with weight $\alpha$ to avoid over-fitting. The objective function forces the normal data features to stay close to the center $c$. Theoretically, the mapping model $\phi$ can be replaced by any neural network architecture, demonstrating the extensibility of our approach. The following two sections show how we develop an appropriate neural network for mapping multiple features. We then describe how to integrate the mapped features for identifying anomalous logs in Section~\ref{sec:integration}.



\subsubsection{Unitary discriminator}
We first look into single logs, as the single log that contains negative words (e.g., ``failure'' and ``error'') usually indicates an anomaly. 
The unitary discriminator works on rich representation of individual logs, aiming to map normal logs to a hyper-sphere that describes the normality.
The motivation behind the unitary discriminator is that, the negative terms in anomalous logs exhibit significantly different semantics than words in normal logs (e.g., ``running'', ``success''). These anomalous logs' features will be mapped far away from the center of the hyper-sphere; thus they are considered as normality-deviating ones.
To this end, we adopt the strong learning ability from neural networks and build the unitary discriminator ($\phi_{uni}$) with a two-layer feed-forward neural network denoted as $FFNN$. We describe the architecture as follows:
\begin{equation}
    \phi_{uni}(x_{rich}; W_{uni}) = FFNN_b((FFNN_a(x_{rich})),
\end{equation}
where $x_{rich}$ refers to the rich representation containing full log semantics (i.e., \textit{unitary feature}) of the center log $x$.

\subsubsection{Local discriminator}

Looking into one individual log is not sufficient to comprehensively understand the running status, so it is noteworthy to exploit its contextual information.
On the one hand, it is pointed out that different logs possess different importance~\cite{zhang2019robust}. For example, some miscellaneous logs regularly appear regardless of what job the system is running, whereas other logs provide richer guidance for analysis. 
On the other hand, log data transmission, collection, and software evolution affect synchronization temporally, leading to unstable sequences.
To focus on beneficial logs and leave the uninformative logs out, we leverage the attention mechanism~\cite{bahdanau2014neural} to focus on beneficial logs.
In the local discriminator, we use the center log and its contexts to acquire a \textit{local feature} against unstable sequences and then learn the normality of such a local feature.

Given a center log $x$, we construct its context representation $x_{ctx}$ by forming its abstract representation of context as a matrix. Then we compute the weights across the context by the attention mechanism~\cite{vaswani2017attention}, allowing the model to learn the importance of surrounding logs, thus addressing the unstable sequence issue.
Specifically, given a center log $x_{rich}$ as query and its context $x_{ctx}$ as value, we compute the weighted context representation as the local feature (denoted as $x_{local}$) as follows:
\begin{equation}
    \begin{aligned}
        x_{local} &= softmax(\frac{x_{query}x_{ctx}^T}{\sqrt{d_k}})x_{ctx},\\
        x_{query} &= FFNN_c(x_{rich}),
    \end{aligned}
\end{equation}
where $d_k$ refers to the dimension of $x_{ctx}$, and $FFNN_c$ transforms $x_{rich}$ to the target domain that shares the same dimension with $x_{ctx}$. 

After that, another two-layer neural network with an activation function is applied to the local feature $x_{local}$ for learning normality from contexts. To sum up, we describe the network for the local discriminator ($\phi_{local}$) as in Equation~\ref{eq:local-discriminator}:
\begin{equation}
    \begin{aligned}
        \phi_{local}(x_{rich},x_{ctx}; W_{local}) = FFNN_e((FFNN_d(x_{local}))).
    \end{aligned}
    \label{eq:local-discriminator}
\end{equation}



\subsubsection{Integration}\label{sec:integration}

The unitary discriminator learns normality for individual logs, whereas the local discriminator learns the context normality in running status. 
To fully exploit these two different information sources, we propose the total objective function with a weighted sum in Equation~\ref{eq:overall-objective} to simultaneously optimize two sub-discriminators: 
\begin{equation}
    \begin{aligned}
        J_{total} = \lambda * J_{uni} + (1-\lambda) * J_{local},
    \end{aligned}
    \label{eq:overall-objective}
\end{equation}
where $J_{uni}$ and $J_{local}$ are the functions defined in Equation~\ref{eq:objctive-func} for unitary discriminator $\phi_{uni}$ and local discriminator $\phi_{local}$, respectively. 
The objective functions allow two discriminators to learn the normality by minimizing their hyper-sphere volume.

The distance between a log message (after mapping by discriminators) to the hyper-sphere center measures the degree of normality of the log.
We apply an \textit{abnormal score} ($abnScore$) to describe how the log deviates from the normality, which is the weighted sum of the abnormal sub-scores from two independent discriminators. 
The abnormal sub-score $abnScore_i$ is defined by the Euclidean distance from the feature embedding to its corresponding hyper-sphere center, denoted by Equation~\ref{eq:abnscore}:
\begin{equation}
    \begin{aligned}
    abnScore &= \lambda * abnScore_{uni} + (1-\lambda) * abnScore_{local} \label{eq:abnscore},\\
    abnScore_i &= \| \phi_i(x; W_i) - c_i \|^2, i \in \{uni, local\}. 
    \end{aligned}
\end{equation}

The center log is eventually predicted as an anomaly if and only if its abnormal score is larger than the threshold $D$ (Equation~\ref{eq:determination}). We put all identified logs into the anomalous log set $AL$, which provides detailed clues to troubleshoot the system conveniently.
\begin{equation}
\mbox{center log} = \begin{cases} \label{eq:determination}
NormalLog, & abnScore \leq D \\
AnomalousLog, & abnScore>D. \\
\end{cases}
\end{equation}

%% file: sections/5Implementation.tex
\section{Implementation Setup}\label{sec:implementation}

\subsection{Data collection}\label{subsec:datacollection}
\subsubsection{Infrastructure}
Despite many log datasets being collected for research~\cite{deeplog2017, loghub, oliner2007supercomputers, Cluster2016}, there is no open-source dataset documenting the evolution process. To fill this blank, we collect a new dataset \textsc{LogEvol} containing log data from the most widely-applied data processing system Spark~\cite{Ean2018MapReduce} (\textsc{LogEvol-Spark}) and Hadoop~\cite{shvachko2010hadoop} (\textsc{LogEvol-Hadoop}), across different versions. 

To this end, we employ HiBench~\cite{hibench}, a big data benchmark suite, to generate logs by running a set of workflows in Spark and Hadoop, respectively, from basic to sophisticated scenarios. 
In total, we run 22 workloads (shown in Table~\ref{tab:workloads}) on the systems to cover more practical scenarios, while other existing datasets~\cite{Cluster2016, loghub} are collected from simply running two straightforward tasks (i.e., page rank and word count).

Then, we repeat the procedure of running workloads using different versions of the software systems mentioned above, covering a wide time range and various data size scales.
We select two typical versions of Spark (i.e., Spark2.4.0 and Spark3.0.3) and Hadoop (i.e., Hadoop2.10.2 and Hadoop3.3.3), as they have undergone systematic changes with significant differences.
\begin{table}[tbp]
\centering
\footnotesize
    \vspace{-0.15in}
    \caption{Workloads for collecting \textsc{LogEvol}.}
    \vspace{-0.1in}
\begin{tabular}{cc}
\toprule
\rowcolor[gray]{.9}
\textbf{Categories} & \textbf{Workloads} \\
\midrule
 Micro task & Sort, Wordcount, etc. \\ 
 Machine learning& Bayes Classification, Gradient Boosted Trees,   etc.\\
 SQL & Aggregation, Join, Scan etc. \\ 
 Websearch & Pagerank \\ 
 Graph & NWeight, Graph Pagerank \\ 
 Streaming & Repartition \\
 \bottomrule
     \vspace{-0.2in}
\end{tabular}
\label{tab:workloads}
\end{table}

\subsubsection{Fault Injection}
We inject 18 typical types of faults into the system to simulate real-world production failures: (1) \textit{Process suspension}: Suspend processes in multiple types of nodes, one at a time; (2) \textit{Process killing}: Kill processes in seven types of nodes, one at a time; (3) \textit{Resource occupation}: Inject other computation programs to occupy CPU and memory; and (4) \textit{Network faults}: Establish network faults such as losing packages, network delay, and connection lost.


In total, we collect 6,703,460 log messages (\# Logs)with recognized 69,513 anomalous logs (\# Anomalous logs), whose statistics are shown in Table~\ref{tab:stats}. To guarantee dataset quality, anomalous logs are discussed and annotated by two engineers who have two-year development experience with the Spark system. Since annotators have read a lot of logs in their development experience, they can provide reliable annotations.

\begin{table}[tbp]
\footnotesize
    \caption{Statistics of \textsc{LogEvol}.}
    \vspace{-0.1in}
    \centering
    \begin{tabular}{c||c|c||c|c}
    \toprule
         &  Spark2 & Spark3 & Hadoop2 & Hadoop3 \\
    \midrule
         \textbf{\# Logs} & 931,960 & 1,600,273 & 2,120,739 & 2,050,488 \\
         \textbf{\# Anomalous logs} & 1,702 & 2,430 & 35,072 & 30,309 \\
    \bottomrule
    \end{tabular}
    \label{tab:stats}
        \vspace{-0.2in}
\end{table}

\subsection{Implementation details}\label{subsec:implementation}

In the multi-level representation extractor, we use BERT~\cite{devlin2018bert} as the pre-trained language model and fine-tune it with Hugging Face~\cite{wolf-etal-2020-transformers}. 
For the anomaly discriminator, we specifically choose leaky ReLU~\cite{xu2015empirical} as the activation function between two layers in the perceptron so as to resolve the ``all-zero-solution" issue~\cite{ruff2018deep}. 
We set the dynamic threshold $D$ to be 0.4 times of the maximum normality (hyper-sphere radius) in the training data in intra-version and 0.6 times for the inter-version.
We set $\lambda$=0.5 in the experiments as the unitary and local features both serve as an important role in fault localization. 
We randomly split the collected logs into training, development, and testing sets for each software version with a standard 8:1:1 splitting. In contrast, the training set only contains logs collected in the fault-free periods as we assume the majority of logs are normal in a healthy system.
All experiments are conducted in 64-bit CentOS 7 with Intel(R) Xeon(R) CPU and 1 GeForce RTX 2080 GPU for acceleration. It takes approximately 15 seconds for the anomaly discriminator to train in an epoch.

%% file: sections/6Experiment.tex
\section{Experiments}\label{sec:exp}






To evaluate the effectiveness of \name, we investigate three research questions: 

\noindent \textbf{RQ1: }How effective is \name~in identifying anomalous logs?

\noindent \textbf{RQ2: }How effective is \name~in resolving evolving events and evolving sequences?

\noindent \textbf{RQ3: }How effective are different components in \name?


\subsection{Experimental settings}
\subsubsection{Baselines}
We select four unsupervised log-based analytics as baselines, including two anomalous log identification models and two anomaly detection (AD) models. LogAnomaly and LogSed are the state-of-the-art AD and log localization models, respectively. The reason why we choose AD baselines is, they both work for anomaly analysis with different granularities (i.e., coarse-grained and fine-grained); 
For AD models, we use the historical sequences to train a reference model and predict the next event as in the original papers. The actual next event that outside the predicted list of candidate events will be considered as anomalous due to its deviation from the reference model. In our implementation, we use the state-of-the-art log parser~\cite{chu2021prefix} to extract events for all baselines as they all require the parsing phase.
In specific, we briefly characterize four baselines as follows.
\begin{itemize}
    \item LOGAN~\cite{tak2016logan} built the diagnosis system by constructing a directed graph from normal log event sequences. Then any of the test time logs that deviate from the directed graph will be considered anomalous.
    \item LogSed~\cite{jia2017logsed} addressed the interleaving logs problem by developing a two-stage approach to mine the important sequential relationship from log sequences. The incoming log message that violates that sequence will be regarded as anomalous.
    \item DeepLog~\cite{deeplog2017} utilizes an LSTM network to capture sequential information of log data. It accepts the sequence of log event IDs to predict the next log, the actual log ID outside prediction will be regarded as an anomaly.
    \item LogAnomaly~\cite{meng2019loganomaly} 
   is proposed for unsupervised anomaly detection with semantic representation for log events via an attention-based LSTM network.
\end{itemize}

\subsubsection{Dataset} \name~is evaluated on two datasets: a software evolution dataset collected from two representative systems~(\textsc{LogEvol}) and a synthetic dataset (\textsc{SynEvol}).

\textbf{\textsc{LogEvol}}. Although existing study~\cite{shang2014exploratory} analyzed the evolution process of Hadoop, and mentioned the importance of new-emerging log messages~\cite{zhang2019robust}, there lacks a public dataset showing how logs change during software evolution. Hence, we evaluate our approach and compare it with baselines on the data collected in Section~\ref{subsec:datacollection}. 
To our best knowledge, \textsc{LogEvol} is the first publicly accessible log dataset recording software evolution activities.

\textbf{\textsc{SynEvol}}. To evaluate how \name~resolves the challenges of unseen events and unstable sequences separately (Note that \name~is parser-free), we build a synthetic dataset based on the collected Spark2 logs in \textsc{LogEvol} (denoted as \textsc{LogEvol}-Spark2). 
Following previous work~\cite{zhang2019robust}, we inject unseen events and unstable sequences into \textsc{LogEvol}-Spark2 to simulate the real-world software evolution as follows:

\textit{1. Unseen events} are introduced by logging statement alteration in software updates. Developers may paraphrase or insert logging statements for customized functionalities. Since \name~does not use a parser, we simulate the change by creating a set of synthetic log messages via (1) inserting, (2) deleting, or (3) replacing a common word from an original log message. 
Such modification is more likely to reflect the changes in log events.

\textit{2. Unstable sequences} occur both in log generation and log evolution. Logs from multiple transaction flows may be interleaving, making the direct predecessor or successor of a certain log different. Moreover, log evolution is likely to cause variations via function ensemble or the changes of function invocation sequences.
To construct synthetic sequences, we randomly remove a few unimportant log messages (far away from anomalous logs), repeat some log messages several times, or shuffle the log messages in a short time. 

We inject the evolving events and unstable sequences into the original dataset, denoted as \textsc{SynEvol}-Events and \textsc{SynEvol}-Seqs correspondingly.
The injection follows specific ratios. We inject the 5\%, 10\%, 15\%, 25\%, and 30\% synthetic log messages and log sequences to \textsc{LogEvol}-Spark2, to observe how \name~reacts to unseen and unstable sequences, respectively. 

\subsubsection{Evaluation metrics}
To evaluate the effectiveness of~\name~in anomalous log identification, we apply Precision, Recall, and F1-score as evaluation metrics. In particular, Precision (P) is the percentage of logs that are correctly identified anomalous overall identified logs ($\frac{TP}{TP+FP}$). Recall (R) is the percentage of logs that are correctly identified anomalous over logs belonging to anomaly logs. ($\frac{TP}{TP+FN}$). F1 score (F1) is the harmonic mean of Precision and Recall ($2*\frac{P*R}{P+R}$),
where $TP$ refers to the amount of anomalous logs that is correctly identified, $FP$ refers to the number of normal logs that are wrongly predicted as anomalous, and $FN$ means the number of anomalous logs that are identified as the normal logs.

\begin{table*}[t]
\footnotesize
    \centering
        \caption{Experimental results in identifying anomalous logs (train set$\rightarrow$test set).}
        \vspace{-0.1in}
    \begin{tabular}{lccc|ccc|ccc|ccc}
    \toprule
    \rowcolor[gray]{.9}
    \multicolumn{13}{c}{\textbf{\textsc{LogEvol-Hadoop}}}\\
    \midrule
    & \multicolumn{6}{c}{\textbf{Intra-version}} & \multicolumn{6}{c}{\textbf{Inter-version}}\\

          & \multicolumn{3}{c}{Hadoop2 $\rightarrow$ Hadoop2} & \multicolumn{3}{c}{Hadoop3 $\rightarrow$ Hadoop3} & \multicolumn{3}{c}{Hadoop2 $\rightarrow$ Hadoop3} & \multicolumn{3}{c}{Hadoop3 $\rightarrow$ Hadoop2}\\
     Baseline & Precision & Recall & F1 & Precision & Recall & F1 & Precision & Recall & F1 & Precision & Recall & F1\\
    \midrule

    LOGAN & 0.894 & 0.995 & 0.942 & 0.899 & 0.988 & 0.942 & 0.360 & 0.988 & 0.528 & 0.376 & 0.995 & 0.546\\
    LogSed & 0.910 & 0.995 & 0.951 & 0.925 & 0.986 & 0.955 & 0.371 & 0.988 & 0.540  & 0.390 & 0.993 & 0.560\\
    DeepLog & 0.913 & 0.985 & 0.947 & 0.926 & 1.000 & 0.961 & 0.386 & 0.999 & 0.556 & 0.410 & 0.971 & 0.576 \\
    LogAnomaly & 0.926 & 0.994 & 0.958  & 0.939 & 0.988 & 0.963 & 0.389 & 0.998 & 0.560 & 0.407 & 0.995 & 0.578 \\
    \textbf{\name} & 0.945 & 0.982 & \textbf{0.963} &  0.952 & 0.988 & \textbf{0.970} & 0.770 & 0.941 & \textbf{0.847} & 0.857 & 0.913 & \textbf{0.884} \\
         \midrule
        \rowcolor[gray]{.9}
    \multicolumn{13}{c}{\textbf{\textsc{LogEvol-Spark}}}\\  
    \midrule
    
     & \multicolumn{6}{c}{\textbf{Intra-version}} & \multicolumn{6}{c}{\textbf{Inter-version}}\\
     & \multicolumn{3}{c}{Spark2 $\rightarrow$ Spark2} & \multicolumn{3}{c}{Spark3 $\rightarrow$ Spark3} & \multicolumn{3}{c}{Spark2 $\rightarrow$ Spark3} & \multicolumn{3}{c}{Spark3 $\rightarrow$ Spark2}\\
     Baseline & Precision & Recall & F1 & Precision & Recall & F1 & Precision & Recall & F1 & Precision & Recall & F1\\
    \midrule

        LOGAN & 0.798 & 0.943 & 0.865 & 0.967 & 0.870 & 0.916 & 0.016 & 0.943 & 0.032 & 0.012 & 0.943 & 0.024\\
    LogSed & 0.842 & 0.914 & 0.877 & 0.907 & 0.923 & 0.915 & 0.013 & 0.917 & 0.026  & 0.010 & 0.914 & 0.020\\
    DeepLog & 0.862 & 0.952 & 0.905 & 0.858 & 0.976 & 0.914 & 0.017 & 0.947 & 0.032 & 0.014 & 0.909 & 0.026 \\
    LogAnomaly & 0.931 & 0.939 & 0.935  & 0.898 & 0.947& \textbf{0.922} & 0.020 & 0.923 & 0.038 & 0.017 & 0.948 & 0.034 \\
    \textbf{\name} & 0.970 & 0.974 & \textbf{0.972} &  0.944 & 0.888 & 0.915 & 0.922 & 0.700 & \textbf{0.795} & 0.920 & 0.812 & \textbf{0.863} \\
         \bottomrule
    
    \end{tabular}
    \label{tab:main-results}
    \vspace{-0.1in}
\end{table*}

\subsection{RQ1: How effective is \name~in identifying anomalous logs?} 
To evaluate how effective \name~can pinpoint the anomalous logs with software evolution activities, we conduct experiments on our dataset \textsc{LogEvol}. The experiments engage two different settings: 1) Intra-version: identify the anomalous logs on the same system it is trained (e.g., Spark2 $\rightarrow$ Spark2); and 2) Inter-version: identify the anomalous logs in a different system version after training (e.g., Spark2 $\rightarrow$ Spark3).

We can draw two observations from the experimental results shown in Table~\ref{tab:main-results}.
First, \name~delivers an overall satisfactory performance under the intra-version setting with the average F1 score of 0.967 in Hadoop and 0.944 in Spark, which is comparable with other baselines. The experimental results indicate that \name~can learn the normality and effectively identify anomalous logs from log sequences. 
Besides, we find that deep learning-based approaches perform better than FSM-based approaches, demonstrating that neural networks are capable of capturing intrinsic sequential patterns and log semantics.


Second, in the inter-version scenario, \name~significantly outperforms all baselines by a wide margin, demonstrating its effectiveness and robustness in software evolution. 
We observe that all baseline performances drastically drop (approximately an F1 score of 0.55) while \name~achieves an average F1 score of 0.87 for Hadoop, which contains 3\% new logs. 
In the case of Spark, where logging statement paraphrasing and insertion via software updating account for 10\% logs, baseline performances are further significantly downgraded.


We analyze the reasons below.
First, log parsers will generate unseen events when they encounter these new logs. 
Then, directed graph approaches (i.e., LOGAN, LogSed) and AD models (i.e., DeepLog, LogAnomaly) fail in matching these unseen events to any current events or predicted subsequent-event candidates.
Consequently, current baselines label all unseen events as anomalous, leading to high false-positive rates (i.e., low precision). 
On the contrary, \name~uses hierarchical clustering to learn abstract representations of log messages and aligns unseen events to similar past ones.
In this way, the modified log message shares the consistent representation with its old one, so as to reduce false positives and improve anomalous log identification performance.
Note that false positive rates in anomalous log identification, although not as severe as false negative cases, can still be problematic as they can lead to excessive work for maintainers.


\begin{tcolorbox}[boxsep=1pt,left=2pt,right=2pt,top=3pt,bottom=2pt,width=\linewidth,colback=white!90!gray,boxrule=0pt, colbacktitle=white!30!gray,toptitle=2pt,bottomtitle=1pt,opacitybacktitle=0.4]
\textbf{Answers to RQ1:} \name~can effectively identify anomalous logs under both intra- and inter-version settings, all the while demonstrating its robustness and stability across software evolution activities.
\end{tcolorbox}



\subsection{RQ2: How effective is \name~in resolving evolving events and evolving sequences?}
\begin{figure}[tpb]
    \centering
	\subfigure[\textsc{SynEvol}-Events example]{
		\includegraphics[width=0.46\linewidth]{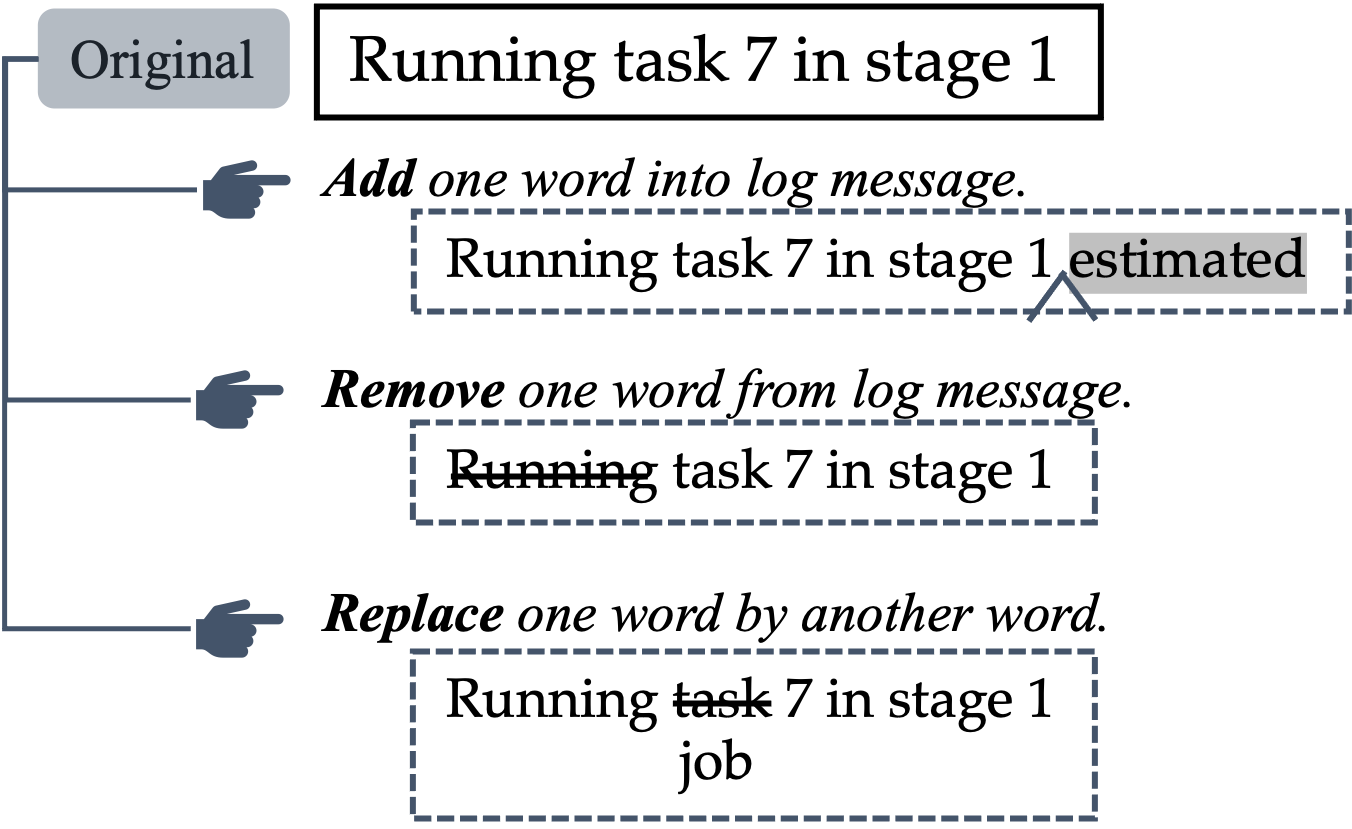}
	}       
	\subfigure[\textsc{SynEvol}-Seqs example]{
		\includegraphics[width=0.46\linewidth]{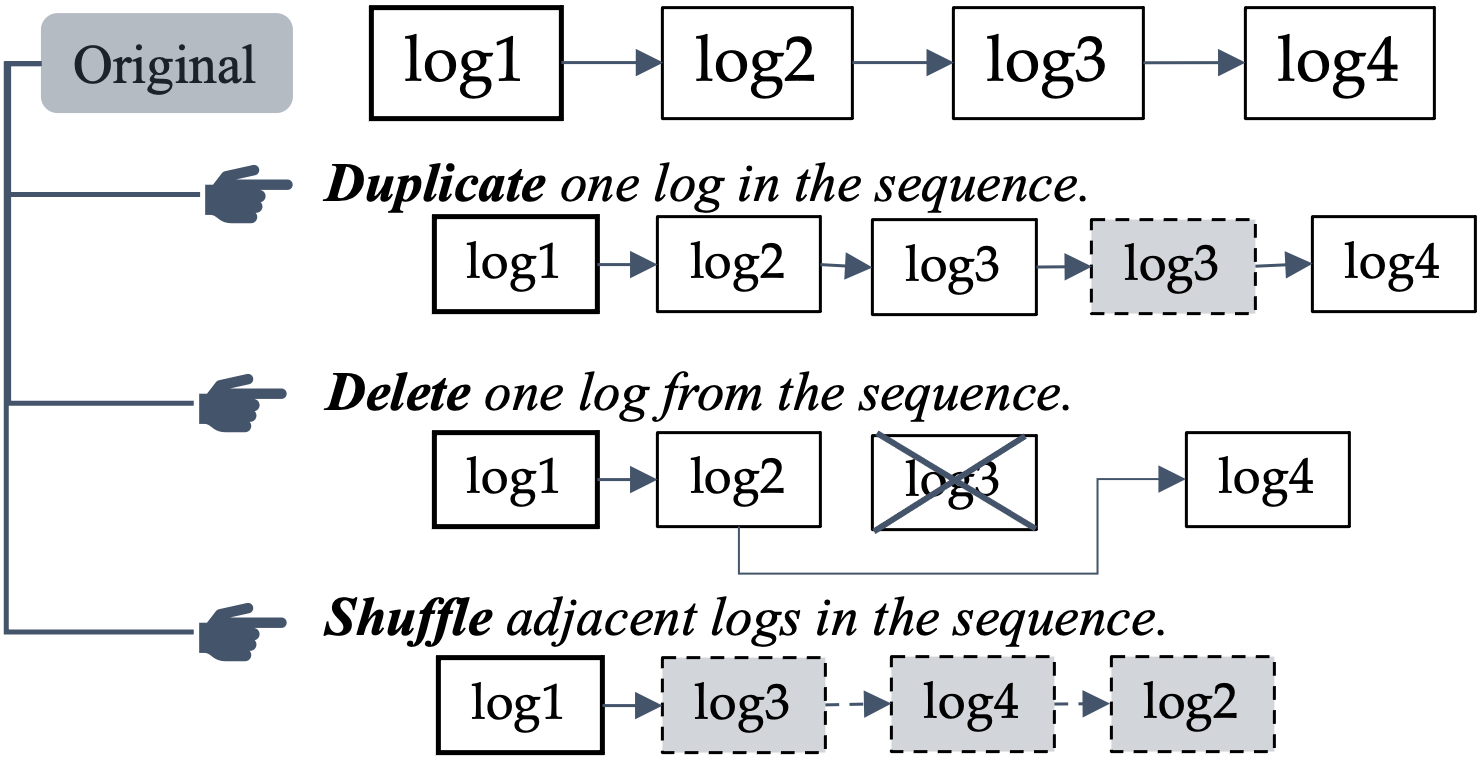}
	} 
 \vspace{-0.1in}
	\caption{Examples of synthetic dataset \textsc{SynEvol}.} 
	\label{fig:syn-example}
\end{figure}

We overcome the parsing errors challenge naturally since our model is parser-free. 
Thus, we are interested in how well our model addresses the other two challenges, i.e., evolving events and evolving sequences. To do so, we measure \name~on the synthetic dataset, including \textsc{SynEvol}-Events and \textsc{SynEvol}-Seqs.
Fig.~\ref{fig:syn-example} shows the examples in the dataset.

\begin{figure}[tpb]
    \centering
	\subfigure[Experiments in \textsc{SynEvol}-Events]{
		\includegraphics[width=0.46\linewidth]{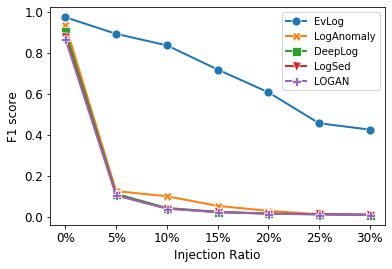}
	}       
	\subfigure[Experiments in \textsc{SynEvol}-Seqs]{
		\includegraphics[width=0.46\linewidth]{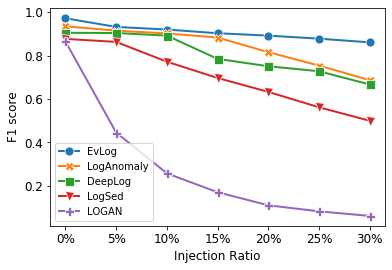}
	} 
 \vspace{-0.1in}
	\caption{Experiment results on the synthetic dataset \textsc{SynEvol}.} 
 	\vspace{-0.15in}
	\label{fig:injection}
\end{figure}

Fig.~\ref{fig:injection} shows the F1 scores of baselines, and ours under the injection ratio varies from 0\% to 30\% (the injection ratio of 30\% means 30\% of the original dataset was replaced by the synthetic one). The results demonstrate our approach's effectiveness in both evolving events and sequences compared with baselines. In particular, \name~achieves the F1 scores of 0.42 and 0.86 in \textsc{SynEvol}-Events and \textsc{SynEvol}-Seqs, even though the synthetic dataset replaces 30\% of the messages and sequences in \textsc{LogEvol}-Spark2, respectively. We attribute the advantage to the extracted multi-level semantics, as well as the stability of the normality learned by the anomaly discriminator.

Another observation is that log changes are more likely to damage the model's performance than sequence changes. This is because log changes bring unseen events to the trained model, posing greater difficulties for the model to deal with. 
On the one hand, our approach can still perform stably with evolving events due to \name's unique clustering mechanism that aligns old events with the new ones. This result is in line with our experiments in RQ1 that all baselines perform unsatisfactorily during version transferring, as many events are changed from Spark2 to Spark3.
On the other hand, in terms of the unstable sequences, we conclude that neural networks (used by LogAnomaly, DeepLog, and ours), particularly those with the attention mechanism (used by LogAnomaly and ours), force the model to pay attention to the informative log messages while getting rid of unstable sequences. 
\begin{tcolorbox}[boxsep=1pt,left=2pt,right=2pt,top=3pt,bottom=2pt,width=\linewidth,colback=white!90!gray,boxrule=0pt, colbacktitle=white!30!gray,toptitle=2pt,bottomtitle=1pt,opacitybacktitle=0.4]
\textbf{Answers to RQ2:} \name~reveals the robustness across different types of changes happening in software evolution, owing to its multi-level semantics extractor and attention mechanism.
\end{tcolorbox}

\subsection{RQ3: How effective are different components in \name?}
This research question investigates an ablation study on how much each design contributes to \name. 
Specifically, we remove each focused component one at a time and conduct experiments on \textsc{LogEvol-Spark}. In particular, we remove (1) the fine-tuning phase in PLM, (2) the unitary discriminator, and (3) the local discriminator, separately. 

\begin{figure}[tpb]

    \centering
	\subfigure[Spark2 $\rightarrow$ Spark2]{
		\includegraphics[width=0.46\linewidth]{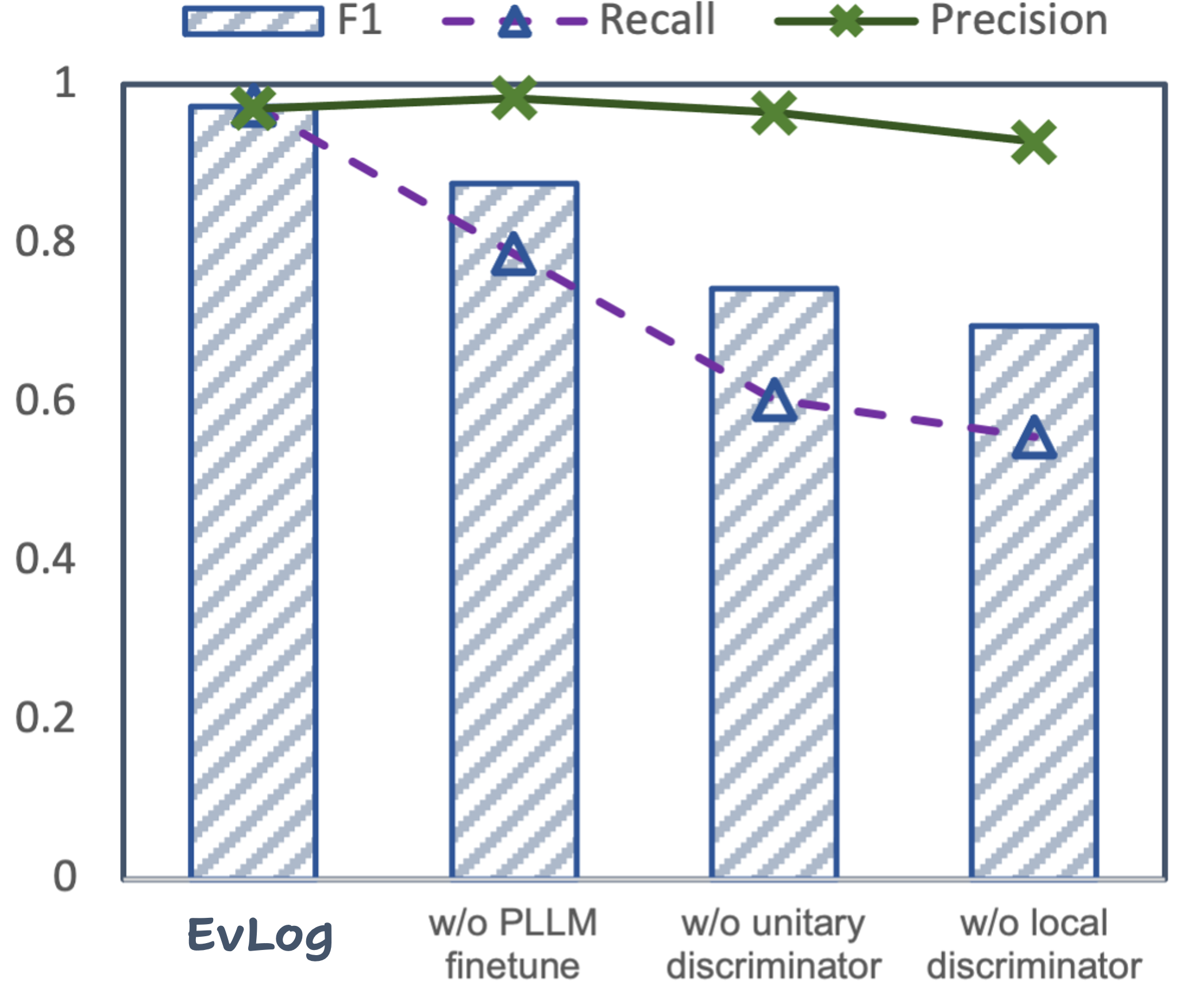}
	}       
	\subfigure[Spark2 $\rightarrow$ Spark3]{
		\includegraphics[width=0.46\linewidth]{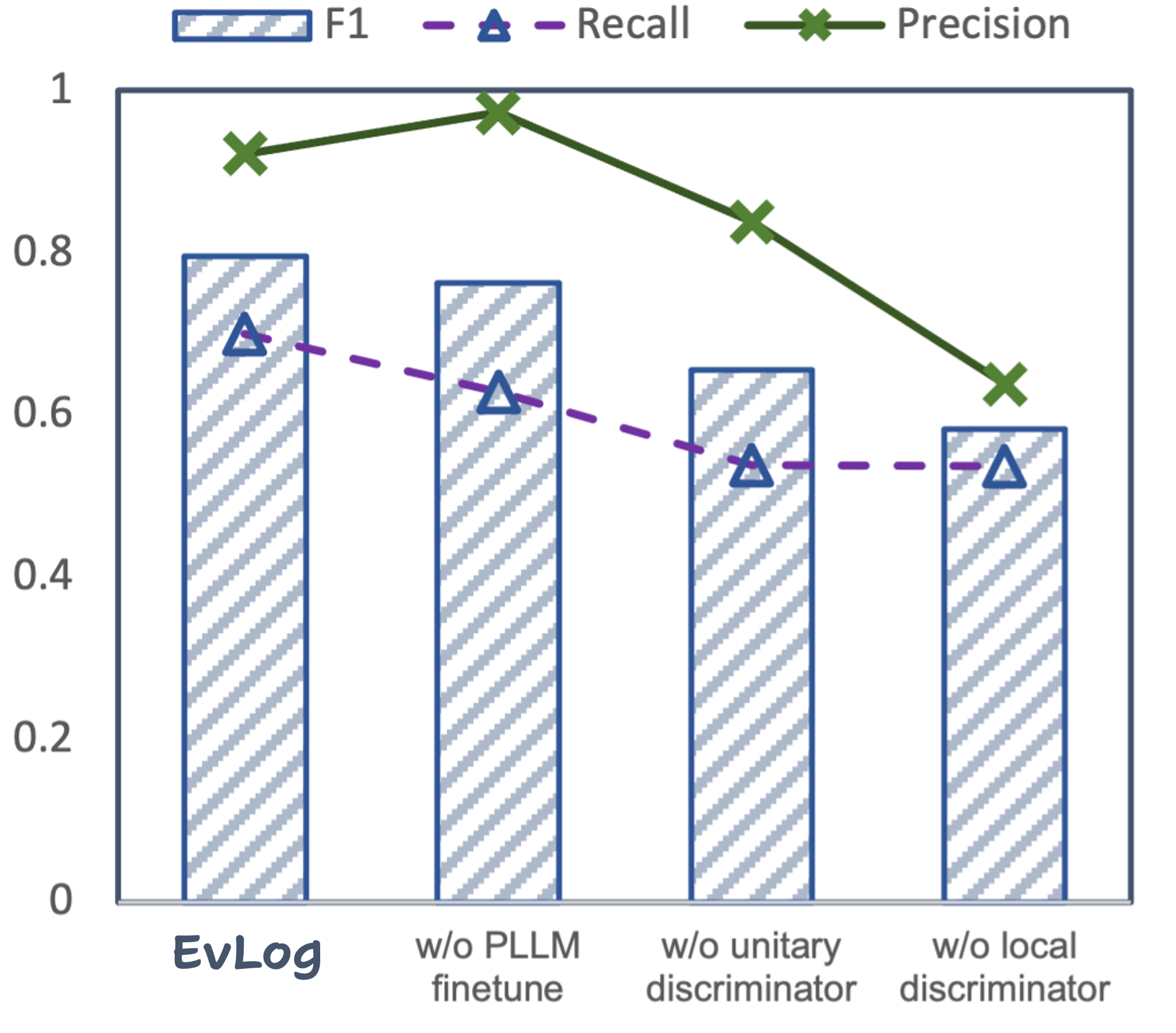}
	} 
 \vspace{-0.1in}
	\caption{Effectiveness of finetuning, unitary discriminator and local discriminator, respectively (train set$\rightarrow$test set).} 
	\vspace{-0.15in}
	\label{fig:ablation}
\end{figure}

Our experiments in Fig.~\ref{fig:ablation} show that all three components of \name contribute to its effectiveness. 
The reasons based on the experiments are elaborated as follows. First, fine-tuning on the log dataset helps \name capture precise semantics by bridging the knowledge gap between Spark domain knowledge and common sense knowledge. 
Second, the unitary discriminator, which operates on individual logs, learns the commonality of single normal logs.
Third, removing the local discriminator largely degrades the overall performance since it provides a more comprehensive view of the contextual running status.

\begin{tcolorbox}[boxsep=1pt,left=2pt,right=2pt,top=3pt,bottom=2pt,width=\linewidth,colback=white!90!gray,boxrule=0pt, colbacktitle=white!30!gray,toptitle=2pt,bottomtitle=1pt,opacitybacktitle=0.4]
\textbf{Answers to RQ3:} The three components, i.e., PLM fine-tuning, unitary discriminator, and local discriminator, all show their effectiveness in the intended design of \name.
\end{tcolorbox}

\begin{figure*}[ht]
    \centering 
    \vspace{-0.15in}
    \includegraphics[width=\textwidth]{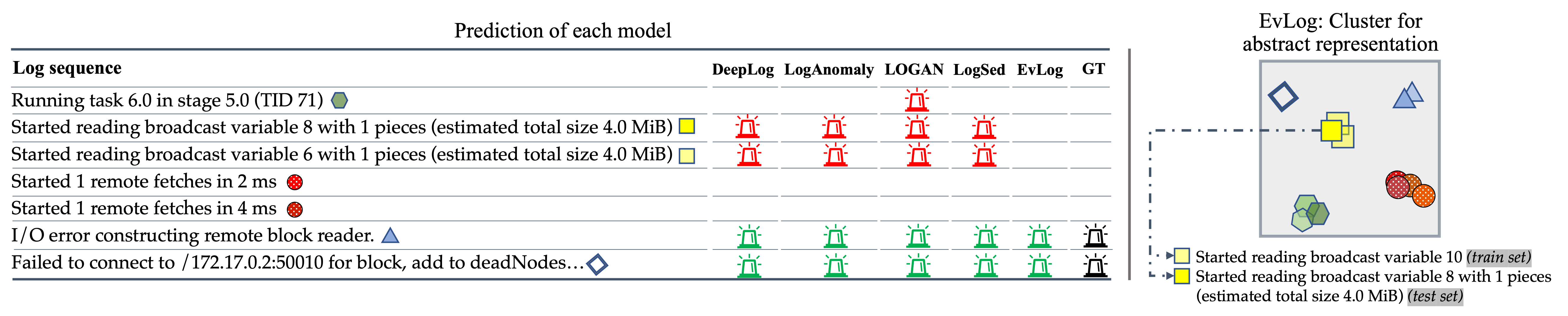} 
    \vspace{-0.35in}
    \caption{An example of how EvLog identifies anomalous logs over software evolution.} 
    \label{fig:case-study} 
    \vspace{-0.2in}
\end{figure*}

%% file: sections/7CaseStudy.tex
\section{Case study}\label{sec:case-study}

This section conducts a case study (Fig.~\ref{fig:case-study}) to show how \name~successfully deals with unseen events and avoids false positives.  
Having been trained on Spark2, baselines and \name~are tested in the case from Spark3, where their $AL$ predictions are marked with lights. Green, red lights refer to true positive and false positive, respectively. ``GT'' refers to the ground-truth $AL$ set. All baselines wrongly predict the line2 and line3 logs as anomalous. We attribute the false-positive results on the two logs to their evolving events. In fact, this event is paraphrased as follows:

\begin{center}
\small
    \fbox
    {\shortstack[l]{
    Spark2: Started reading broadcast variable $<$*$>$ \\
    \rule[0.05\baselineskip]{0.45\textwidth}{0.5pt}\\
    Spark3: Started reading broadcast variable $<$*$>$ with 
    $<$*$>$ pie-\\ces (estimated total size $<$*$>$ MiB)}
    }
\end{center}
where $<$*$>$ refers to the run-time generated numeric values.


Facing such evolving logs, \name~can mitigate the associated issue by the abstract representation shown on the right-hand side of Figure~\ref{fig:case-study}. Though line2 and line3 are unseen logs, they can be assigned to a cluster that contains historical semantically similar log messages, according to their rich semantic representations. 
The yellow squares represent the rich representations of logs in the hyperspace, where the logs before and after paraphrasing stay closely in one cluster. Therefore, the high-level abstract representation remains stable in the change from the original logs to the paraphrased logs, and these new logs will not be mapped far away from the hyper-sphere's center. Eventually, the model can identify the new paraphrased log as a normal one because it does not deviate from the normality.
\vspace{-0.05in}

%% file: sections/8ThreatLiterature.tex
\section{Threat to Validity}\label{sec:threats}

\textbf{Internal threats.}
(1) Dynamic threshold. \name~requires a dynamic threshold $D$ to identify anomalous logs. 
Our study found that the satisfied threshold for intra-version and inter-version identification is 0.4 and 0.6 times the maximum normality in the training data, respectively.
The threshold strikes a balance between recall rate and precision rate.
In practice, maintainers can customize the threshold based on different scenarios.
(2) Domain knowledge gap.
Technical terms in logs may have specific meanings not captured by PLMs. For example,  we use ``volume'' to describe a detachable block storage device in a computing system, but it usually refers to the degree of loudness or the amount of space in daily life. We fine-tune the PLM with the collected log messages to mitigate the threat.

\textbf{External threats.}
(1) Software drastic evolution. Software systems possibly experience a drastic change, such as complete code restructuring or infrastructure renewal. In such scenarios, logging statements are likely to be altered significantly, and our approach has limitations to handle it without incremental learning. 
Nevertheless, our comparison between Spark2 and Spark3 over two years shows limited extreme changes.
(2) Limited dataset. \name~has been evaluated with only two real datasets and a synthetic dataset, and more real datasets with diverse job types are necessary to validate \name's effectiveness.
However, as this is a brand-new task, datasets are sparse and challenging to collect. To address this issue, our created dataset is collected with representative 22 benchmark workloads from two widely-used systems. 
Although this dataset does not cover all possible workloads, it includes many commonly used ones and provides a practical simulation of the task.

\section{Related Work}\label{sec:related-work}
\subsection{Software evolution}
Run-time data of systems can vary dramatically from time to time, as cloud systems are continuously being upgraded and evolving, causing variations in statistical properties \cite{Stability}. Some studies \cite{shang2014exploratory, zhang2019robust, Stability, LogTracker} noticed this issue and conducted empirical studies to investigate the effects it brings to automated techniques and found many methods are not intelligent enough to embrace such evolution.
For example, previous research~\cite{shang2014exploratory} found that the frequent source codes change liking releasing a new version leads to fragile log processing techniques. Also, LogTracker~\cite{LogTracker} revealed that it is challenging to request developers to maintain well-organized log statements as software evolves without rigorous specifications and demonstrated that the vast majority of context-similar logs come from log reversions. 

These studies demonstrate that software evolution poses challenges to automated log analytics. Yet, they have not directly pointed out how and to what degree such an issue will affect log analytical tools. This paper is the first systematic study that fills such a gap by pointing out the three challenges log evolution brings and their reasons, as well as proposing a solution to overcome software evolution issues.

\subsection{Failure analysis}
Tremendous efforts have been devoted to cloud reliability insurance, and failure analysis has attracted considerable attention since it provides detailed clues for troubleshooting.
Some existing approaches look deep into the source code to localize the failures, for example, mapping log messages to source code and reconstructing execution process for debugging~\cite{chen2019empirical}.
However, source codes are not always accessible. Recently, log-based failure analysis is in the ascendant.
To highlight the anomalous logs for failure diagnosis, existing approaches attempt to abstract the state transition processes in normal status by mining the logs and identifying the log that deviated from the model.
For example, previous studies~\cite{tak2016logan, babenko2009ava} built a directed graph by regrading a log message sequence as an execution workflow and then checked whether logs in the test phase deviate from the graph.
Besides, some studies use a retrieval-based approach to map a newly identified failure into the historical failure database whose cause is annotated by an expert in advance~\cite{jiang2017causes, amar2019mining}. 

However, these methods share three shortcomings in the evolution scenario. First, they rely strictly on dependencies within historical data. Second, they cannot extract sufficient semantics from logs, which are found significant in software evolution. Third, these approaches highly rely on prior expertise, making it impractical in modern evolving systems.

%% file: sections/9Conclusion.tex
\section{Conclusion}\label{sec:conclusion}

Existing advanced log localization models are proposed to discover anomalous logs that may indicate faults in a system automatically, but they ignore software evolution activities. This paper first empirically identifies three challenges (i.e., parser errors, evolving events, and unstable sequences) carried with software evolution and discusses how these challenges can affect localization models. 
Second, we propose~\name~to address the above challenges. To deal with the first two challenges, we develop a parser-free extractor to mine multi-level semantic representation from logs. Then, an anomaly discriminator with an attention mechanism is built to overcome the unstable sequence issue. 
At last, the effectiveness of \name~in identifying anomalous logs over software evolution is confirmed by evaluating it on large-scale system logs. 
This is a newly identified research task in anomalous log localization due to software evolution, and the associated code of \name~as well as the newly collected datasets, are released for research purposes.
We hope our study can motivate more future work on software evolution in the log analytics community.


\section{Acknowledgement}

The work described in this paper was supported by the National Natural Science Foundation of China (No. 62202511), the Key-Area Research and Development Program of Guangdong Province (No. 2020B010165002), and the Key Program of Fundamental Research from Shenzhen Science and Technology Innovation Commission (No. JCYJ20200109113403826).